\documentclass[%
 reprint, nofootinbib,
 amsmath,amssymb,
 aps,]{revtex4-1}

\usepackage{graphicx}
\usepackage{dcolumn}
\usepackage{bm}
\usepackage{float}
\usepackage{placeins}
\usepackage{subfigure}
\usepackage{caption}
\begin{document}


\title{Zero-Point Energies prevent a Trigonal to Simple Cubic Transition in High-Pressure Sulfur}

\author{Jack Whaley-Baldwin}
 \email{jajw4@cam.ac.uk}
 \affiliation{TCM Group, Cavendish Laboratory, University of Cambridge\\}

\date{\today}

\begin{abstract}
Recently published \cite{Whaley_Baldwin} Density Functional Theory results using the PBE functional suggest that elemental sulfur does not adopt the simple-cubic (SC) $Pm\bar{3}m$ phase at high pressures, in disagreement with previous works \cite{Rudin,USPEX}. We carry out an extensive set of calculations using a variety of different functionals, pseudopotentials and the all-electron code \verb|ELK|, and we are now able to show that even though under LDA and PW91 a high-pressure simple-cubic phase does indeed become favourable at the static lattice level, when zero-point energies (ZPEs) are included, the transition to the simple-cubic phase is suppressed in every case, owing to the larger ZPE of the SC phase. We reproduce these findings with pseudopotentials that explicitly include deep core and semicore states, and show that even at these high pressures, only the $n=3$ valence shell contributes to bonding in sulfur. We further show that the $Pm\bar{3}m$ phase becomes even more unfavourable at finite temperatures. We finally investigate whether anharmonic corrections to the zero-point energies could make the $Pm\bar{3}m$ phase favourable, and find that these corrections are several orders of magnitude smaller than the ZPEs and are thus negligable. These results therefore confirm the original findings of \cite{Whaley_Baldwin}; that the high pressure transition sequence of sulfur is $R\bar{3}m \rightarrow$ BCC, with no intervening SC phase.
\end{abstract}

\maketitle

\section{\label{intro}Introduction}

High-pressure first-principles investigations on sulfur \cite{Whaley_Baldwin,Rudin,USPEX} have identified single-atom simple-cubic $Pm\bar{3}m$ sulfur as an energetically competitive phase in the range $300-500$ GPa, where it competes with a single-atom trigonal $R\bar{3}m$ phase. Using the LDA \cite{Rudin}, and PBE \cite{USPEX} exchange-correlation functionals, previous authors have concluded that a transition to the $Pm\bar{3}m$ phase occurs at around $280$ GPa, whereas \cite{Whaley_Baldwin} finds (using PBE) that the simple-cubic phase is not favourable at either the static-lattice or ZPE-included level of theory. All three authors are however in agreement that sulfur eventually adopts a primitive BCC $Im\bar{3}m$ structure at a pressure of around $500$ GPa.

\begin{figure}[!htbp]
\hspace{0.075cm}
\subfigure{    \includegraphics[width=0.17\textwidth]{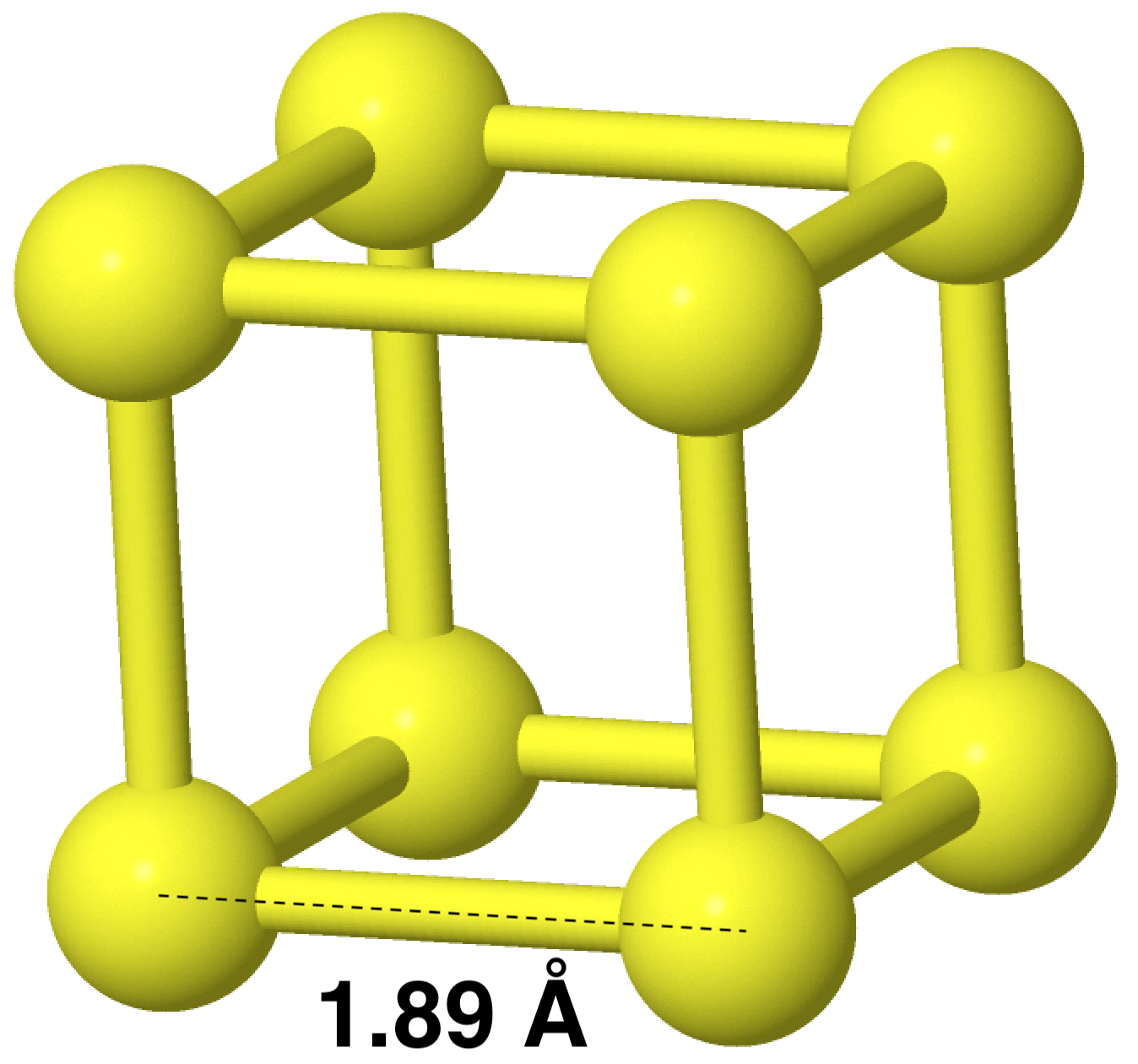}}
\hspace{0.525cm}
\subfigure{\includegraphics[width=0.21\textwidth]{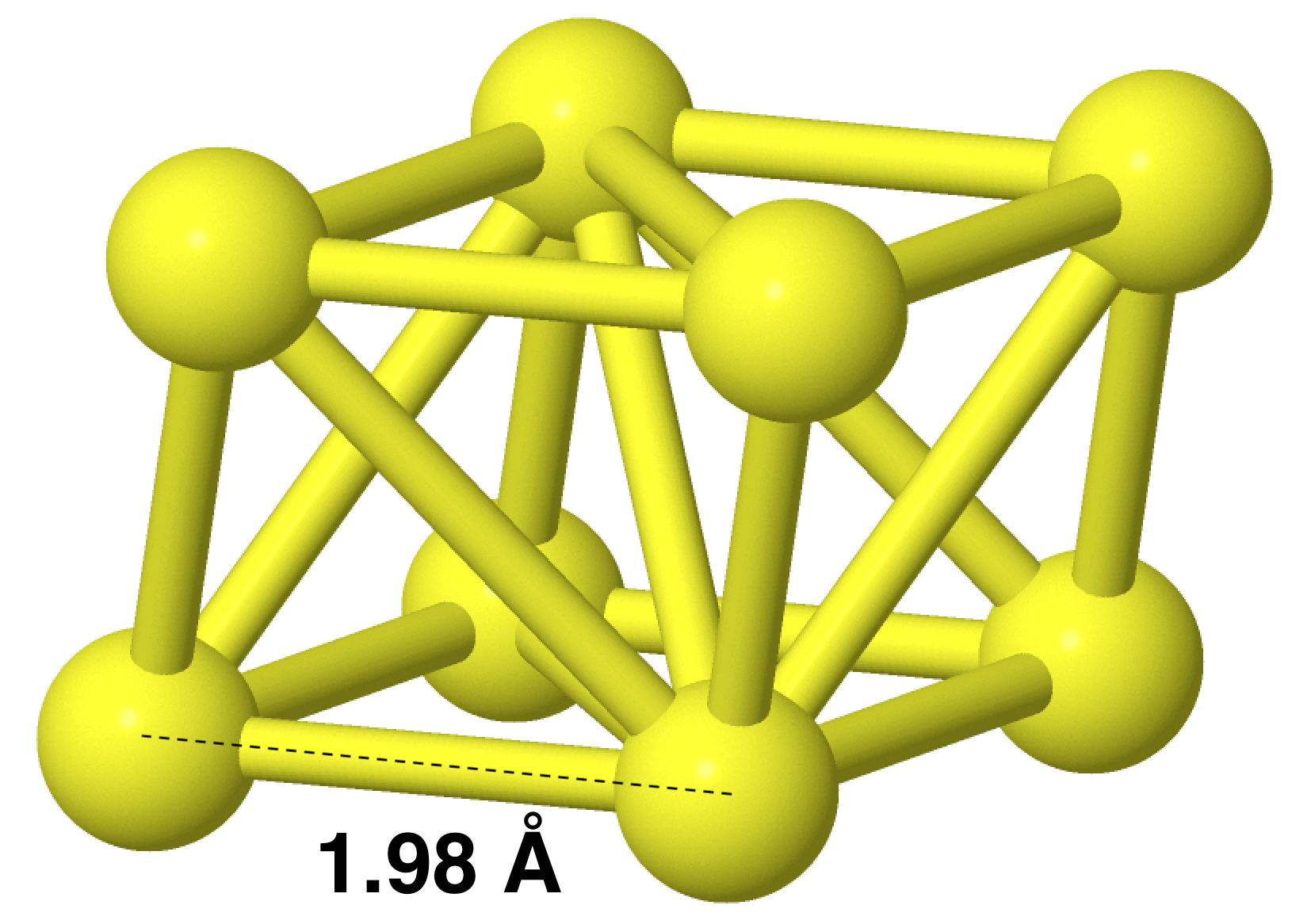}}
\hfill
\caption{The primitive $Pm\bar{3}m$ simple cubic (left) and primitive $R\bar{3}m$ trigonal structures of sulfur at $350$ GPa according to the LDA. The rhombohedral lattice angle for the trigonal structure is $104.9 ^{\circ}$}
\label{structures}
\end{figure}

The existence of a simple-cubic phase in sulfur would be significant, as the $Pm\bar{3}m$ space group is, even at high pressures, only rarely encountered among elemental crystal structures as a lowest-enthalpy phase, primarily due to its highly unfavourable packing efficiency.

\begin{figure*}[!htbp]
    \hspace{-0.85cm}
    \includegraphics[width=1\textwidth]{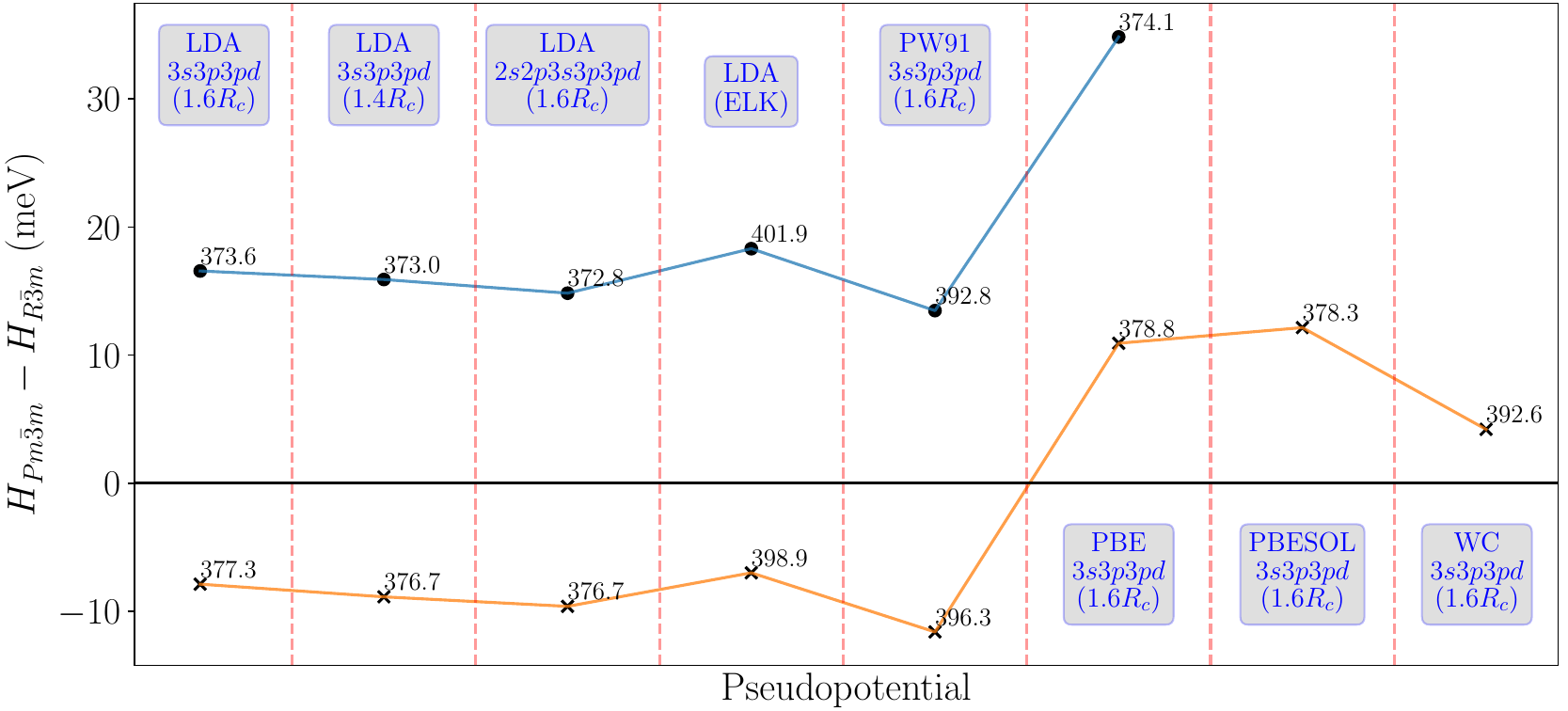}
    \caption{Enthalpy of the simple-cubic $Pm\bar{3}m$ phase of sulfur relative to the trigonal $R\bar{3}m$ phase using different functionals and pseudopotentials. Crosses denote static lattice enthalpies, and filled-in circles denote static lattice enthalpies with zero-point energies included. For the PBESOL and WC cases, the zero point energies were not calculated. The pressure at which the simple-cubic phase is lowest in energy relative to the $R\bar{3}m$ phase is shown next to each datapoint. The lines are a guide to the eye. NOTE: The ELK code does not use pseudopotentials, but rather is an all-electron code.}.
    \label{SC_vs_trigonal}
\end{figure*}

\section{\label{comp_details}Computational Details}

We used the plane-wave DFT code \verb|CASTEP| \cite{Castep} for our electronic structure calculations. A Fermi-Dirac electronic smearing temperature of $500$ K and a $k$-point spacing of $0.01$\AA$^{-1}$ (giving $\approx 3000$ kpoints in $Pm\bar{3}m$ sulfur at $350$ GPa) was used throughout. All geometry optimisations were carried out such that the force on each atom was less than $1 \times 10^{-5}$ eV\AA$^{-1}$, and the differential stress on each unit cell less than $1 \times 10^{-4}$ GPa.
\\\\
In this study, we considered a variety of different pseudopotentials and XC-functionals, which are detailed in table \ref{pseudo_table}.
The plane-wave cutoff values specified in the table were sufficient to converge the absolute energy of each pseudopotential to better than $\pm 0.5$ meV, with the relative convergence between structures being even better than this (see supplemental material).

\begin{table}[!htbp]
\begin{tabular}{|l|l|l|l|}
\hline
Projectors \& Functional \ & $R_c$ (Bohr) \ & $l_{loc}$ \ & Cutoff (eV) \ \\ \hline
$3s3p3d$, LDA & $1.6$  & $3$  & $800$ \\ \hline
$3s3p3d$, LDA       & $1.4$  & $3$  &  $1000$ \\ \hline
$2s2p3s3p3d$, LDA   & $1.6$  & $3$  &  $1000$\\ \hline
$3s3p3d$, PW91 & $1.6$  & $3$  & $800$ \\ \hline
$3s3p3d$, PBE & $1.6$  & $3$  & $800$ \\ \hline
$3s3p3d$, PBESOL & $1.6$  & $3$  & $800$ \\ \hline
$3s3p3d$, WC91 & $1.6$  & $3$  & $800$ \\
\hline
\end{tabular}
\caption{Details of the different pseudopotentials used in this study. $R_c$ is the cutoff radius for each pseudopotential (chosen to be the same for all projectors) and $l_{loc}$ is the local channel.}
\label{pseudo_table}
\end{table}

The study of high-pressure phases of matter with codes that employ pseudopotentials require careful consideration. Most notably, it is essential that (i) The cutoff radius of the pseudopotentials, when doubled, is appreciably smaller than the smallest bond length(s) and (ii) The correct number of valence electrons are included, which may require, in the definition of the pseudopotential, the inclusion of semi-core states. Including projections onto higher-energy states (which, although unoccupied in the isolated atom, become (partially) occupied in the crystal) may also be important.
\\\\
We used ultrasoft pseudopotentials throughout, created using the \verb|CASTEP| on-the-fly-generation (OTFG) program which is bundled with the code. We used two ultrasoft projectors per orbital in all of the pseudopotential definitions, except for the $2s2p3s3p3d$ pseudopotential, where only one projector was used for each of the $2s$ and $2p$ orbitals. All of the cutoff radii in table \ref{pseudo_table} are less than half the smallest interatomic separation in this study ($\approx 1.82$ \AA \ in SC sulfur at $525$ GPa).
\\\\
The choice of pseudopotentials in table \ref{pseudo_table} reflects a variety of XC-functionals. For the LDA case only, we constructed two extra pseudopotentials: One with an especially small cutoff radius ($1.4$ Bohr), and another that explicitly included the core $n=2$ states as valence; in order to check whether the $R_c$ was small enough, and whether the $n=2$ shell contributes to bonding, respectively. As will be discussed later, and as demonstrated in figure \ref{SC_vs_trigonal}, the smaller cutoff radius and inclusion of the $n=2$ shell actually had a negligible effect on our results.
\\\\
We additionally compare our \verb|CASTEP| LDA results to those of the all-electron code \verb|ELK| \cite{ELK}, which utilises the Full-Potential Linearised Augmented Plane Wave (FLAPW) method. A fixed muffin-tin radius of $1.788$ Bohr was used for the \verb|ELK| calculations and the basis set size parameter $R_{MT} \times$ max $\{ \vec{G},\vec{k}\}=9$, with a $50\times50\times50 \ k$-point grid used throughout. The interstitial density was expanded with $|\vec{G}_{max}|=14$ Bohr$^{-1}$. Since the \verb|ELK| code cannot presently perform non-zero externally applied stress geometry optimisations, the output geometry of the $2s2p3s3p3d$ CASTEP calculation (see table \ref{pseudo_table}) was fed into the \verb|ELK| calculation, and \verb|ELK| was simply used to evaluate total energies.
\\\\
Harmonic phonon calculations were carried out using the finite-differences method within the \verb|CASTEP| code and Density Functional Perturbation Theory (DFPT) within the \verb|ELK| code. A $4\times4\times4 \ q$-point grid was used in both cases.
\\\\
The final total energy $E_{tot} \equiv E_{elec} + E_{phonon}$ and volume values from each optimisation were taken and fitted to the Vinet equation of state \cite{Vinet}. The enthalpy was then derived from the derivative of this curve as $H=E-V\big( \frac{\partial E}{\partial V} \big)_T$ . This approach therefore includes the (harmonic) phonon contribution to the total pressure.

\section{\label{energetic_comparison_SC_trigonal}Energies of the Trigonal and Simple Cubic Phases}

\begin{figure}[!htbp]
    \includegraphics[width=0.5\textwidth]{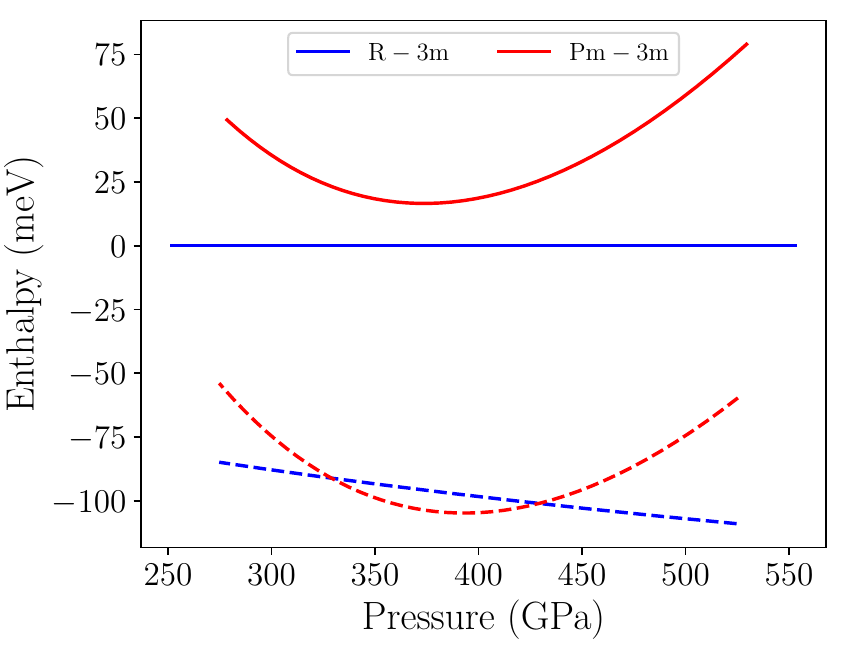}
    
    \vspace{0.225cm}
    
    \includegraphics[width=0.5\textwidth]{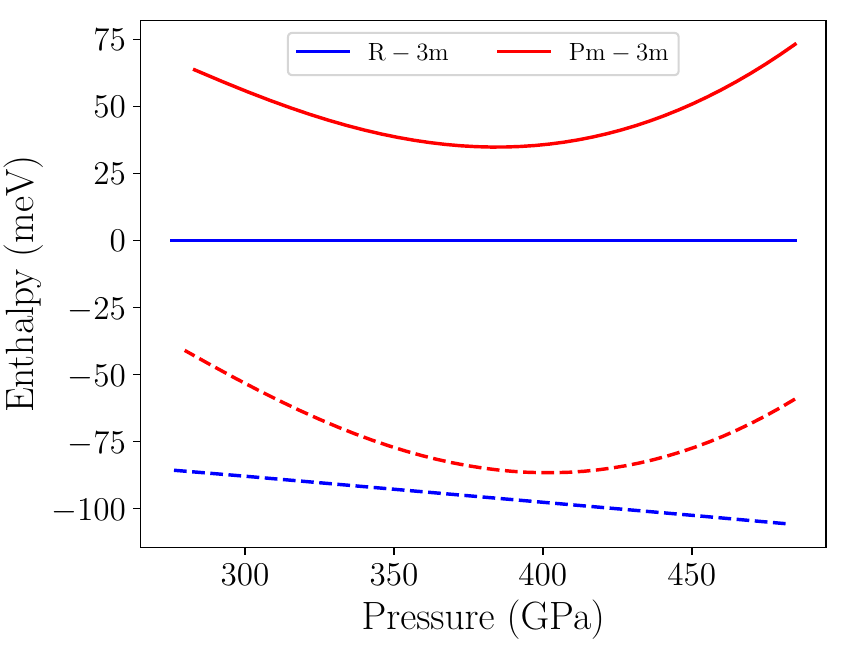}
    \caption{Relative enthalpy curves for the simple-cubic and trigonal phases of sulfur using the LDA functional (top) and PBE functional (bottom). Dotted lines denote energies without ZPEs, solid lines are with ZPEs included. The PW91 functional enthalpy curve is very similar to the top plot, and the PBESOL and WC enthalpy curves are very similar to the bottom plot.}
    \label{gibbs_curves}
\end{figure}

Figure \ref{gibbs_curves} shows relative enthalpy curves (at $0$K) for the $Pm\bar{3}m$ and $R\bar{3}m$ phases of sulfur using the LDA and PBE exchange-correlation functionals, both with and without the inclusion of zero-point energies (ZPEs). Within the LDA, a transition from the $R\bar{3}m$ to the $Pm\bar{3}m$ phase occurs at the static lattice level, but this transition is suppressed when zero-point energies (ZPEs) are included. Using PBE, no transition occurs either with or without ZPEs. The PW91 curve has a shape identical to that of the LDA case with slightly shifted values, and likewise the PBESOL and WC curves have shapes identical to that of the PBE curve with shifted values.

Figure \ref{SC_vs_trigonal} shows the smallest calculated enthalpy of the simple-cubic phase relative to the trigonal phase for a variety of pseudopotentials and exchange-correlation functionals. It can be seen, using LDA and PW91, that on average the simple-cubic phase is lower in energy at the static lattice level by around $9$ meV. On the other hand, PBE, PBESOL and WC have that the simple-cubic phase is higher in energy at the static lattice level, by around $10$ meV for the PBE and PBESOL cases, and around $4$ meV for the WC case. When zero-point energies (ZPEs) are included however, the enthalpy of the simple-cubic phase relative to the $R\bar{3}m$ phase is higher in every single case by at least $12$ meV, rising to $35$ meV for the PBE case.

\begin{figure}[!htbp]
    \hspace{-1.2cm}
    \includegraphics[width=0.5\textwidth]{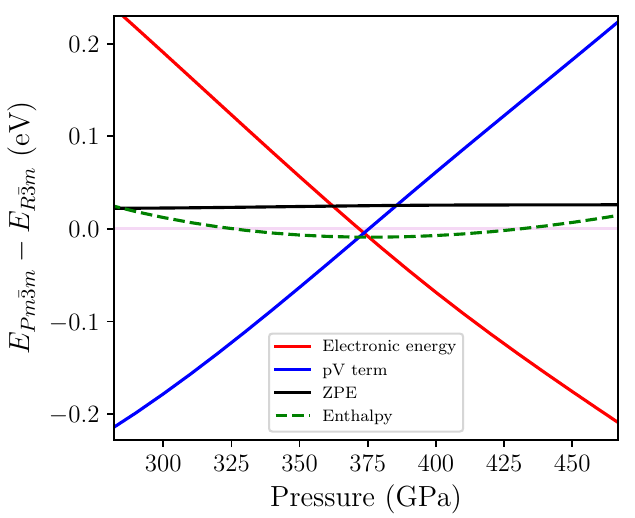}
    \caption{$0$ K electronic energy, the $pV$ term, phonon zero-point-energy (ZPE) and static-lattice enthalpy $H$ of the $Pm\bar{3}m$ phase relative to the $R\bar{3}m$ phase using LDA. Data was taken in $25$ GPa intervals and fitted to a cubic spline. A faint horizontal line at $E=0$ has been drawn for clarity.}
    \label{energy_pv_zpe_enth}
\end{figure}

Figure \ref{energy_pv_zpe_enth} provides further insight by decomposing the energetic contributions to the overall enthalpy of each phase within the LDA. It shows that the $Pm\bar{3}m$ phase becomes energetically favourable at the static lattice level because, below $375$ GPa, its electronic energy falls more quickly than the increase in its $pV$ term. This results in a relative static-lattice enthalpy reduction that reaches a maximum of $\approx 9$ meV at $375$ GPa, with the benefit reducing above this pressure. The zero-point energy (ZPE) of the $Pm\bar{3}m$ phase is, however, consistently higher than that of the $R\bar{3}m$ phase throughout; slowly increasing from $22.3$ meV to $26.1$ meV between $280$ and $470$ GPa respectively. It is the failure of the static-lattice enthalpy reduction to offset the large difference in ZPE that prevents the $Pm\bar{3}m$ from becoming the ground state of sulfur at these pressures.

\begin{figure}[!htbp]
    \includegraphics[width=0.5\textwidth]{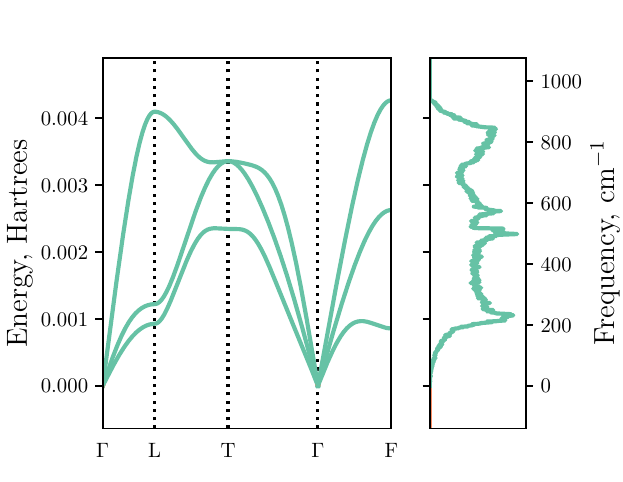}
    
    \vspace{-0.25cm}
    
    \includegraphics[width=0.5\textwidth]{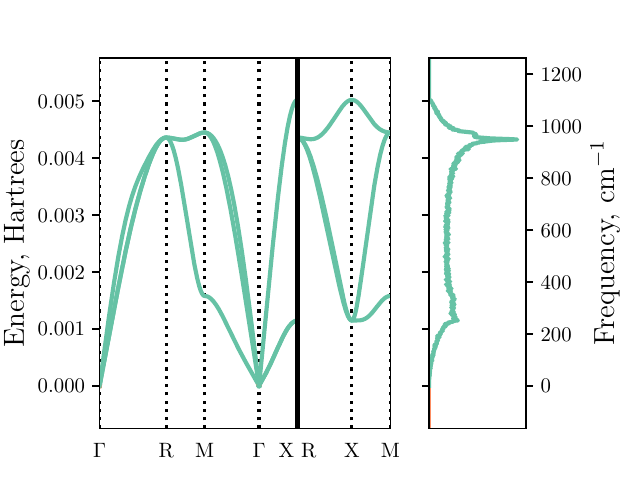}
    
    
    \caption{Phonon dispersion curves for the trigonal $R\bar{3}m$ phase of sulfur (top) and the simple-cubic $Pm\bar{3}m$ phase (bottom) at $375$ GPa using LDA. The high-symmetry points are labelled according to standard convention for the (primitive) trigonal and cubic cells, respectively. The phonon density of states is shown in a panel on the right of each plot.}
    \label{phonon_dispersions}
\end{figure}

Figure \ref{phonon_dispersions} reveals why the ZPE of the SC phase is significantly higher than that of the trigonal phase. Whereas the $R\bar{3}m$ phonon density of states (DOS) plot is shaped approximately like a uniform top-hat function between $200$ and $900$ cm$^{-1}$, the cubic $Pm\bar{3}m$ DOS curve has most of its weight above $600$ cm$^{-1}$, with a pronounced peak at $\approx 950$ cm$^{-1}$. The $Pm\bar{3}m$ DOS curve also extends to higher energies than the $R\bar{3}m$ curve. These result in a lower zero-point energy, which is given by the integral:

\begin{equation}\label{ZPE_integral}
    E_{ZPE} = \frac{1}{2} \int \omega \ g(\omega) \ d\omega
\end{equation}

Where $g(\omega)$ is the phonon density of states. As first pointed out by \cite{Rudin}, the hardening of the phonon modes in the $Pm\bar{3}m$ phase is primarily a result of the shorter bond lengths in this structure (see figure \ref{structures}).

\begin{figure}[!htbp]
    \hspace{-0.6cm}
    \includegraphics[width=0.475\textwidth]{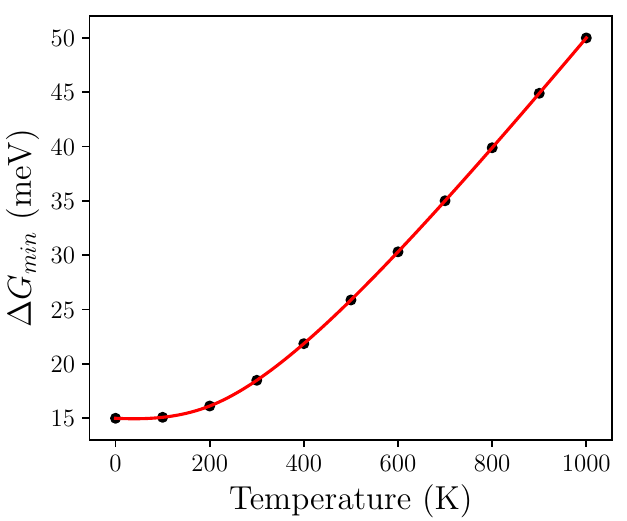}
    \caption{The minimum Gibbs free energy gap in meV, \\ $\Delta G_{min}(T) \equiv G_{Pm\bar{3}m}(T)-G_{R\bar{3}m}(T)$, between the simple cubic and trigonal phases within the harmonic approximation using the LDA functional, plotted as a function of temperature.}
    \label{gibbs_gap_vs_T}
\end{figure}

At finite temperatures, where the quantity determining phase stability is the Gibbs free energy $G(p,T)$, the stability of the $R\bar{3}m$ phase is enhanced, as the minimum Gibbs free energy gap between the phases (occurring at around $375$ GPa with LDA) increases from its $0$ K value of $15$ meV up to $50$ meV at $1000$ K under the harmonic approximation (the \textit{quasi}-harmonic approximation (QHA) was not employed). This is because the phonon free energy $F_{ph}(T,V)$ for the $Pm\bar{3}m$ phase rises more rapily than that of the $R\bar{3}m$ phase. This behaviour is shown in Figure \ref{gibbs_gap_vs_T}. Using the PBE, PBESOL or WC functionals, the Gibbs free energy gap at any given temperature is even larger.

\section{\label{anharmonic}Anharmonic Corrections to Zero-Point Energies}

Having established that the trigonal $\rightarrow$ simple cubic transition does not occur for any XC-functional at the harmonic level, we considered whether anharmonic corrections to the zero-point energies of the $Pm\bar{3}m$ and $R\bar{3}m$ structures would result in a transition with the LDA functional. This would occur if the anharmonic corrections were such that the large energy difference between the ZPEs was reduced significantly.
\\\\
We first computed the harmonic phonons on a $6 \times 6 \times 6$ $q$-point grid, and then used a coarser $2 \times 2 \times 2$ $q$-point grid for the anharmonic calculations. Corrections up to quartic order in the vibrational Hamiltonian were considered, but we did not include any cross-coupling terms between non-degenerate phonon modes. This means that the vibrational Hamiltonian contained terms such as $q_1^3$, $q_2^3$, $q_1^4$, $q_2^4$, $q_1^3q_2$, $q_1^2q_2^2$ etc (where the $q_i$ are phonon normal mode coordinates, and modes $1$ and $2$ are degenerate), but not, for example, $q_1^3q_3$ if modes $1$ and $3$ are not degenerate.
\\\\
It was found that at $375$ GPa (the pressure at which the enthalpy gap between the phases is lowest under LDA) the ZPE of the $Pm\bar{3}m$ structure decreases by $0.272$ meV, and that of the $R\bar{3}m$ structure decreases by $0.218$ meV. This gives an overall relative ZPE change of $0.054$ meV, which is several orders of magnitude smaller than the minimum enthalpy gap between the phases at the harmonic level (15 meV). Therefore, the inclusion of anharmonic corrections to the ZPEs does not change our conclusions.
\\\\
These findings also demonstrate that sulfur is strongly harmonic even at these large pressures.

\section{\label{elec_DOS}Electronic Densities of States}

\begin{figure}[!htbp]

    \hspace{-0.4cm}
    \includegraphics[width=0.5\textwidth]{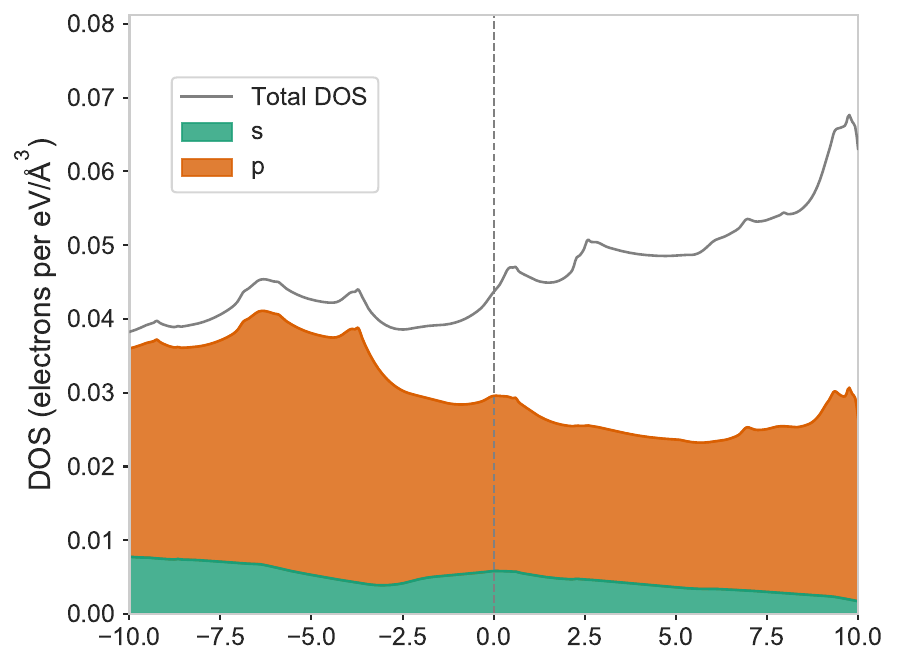}

    \vspace{0.25cm}

    \hspace{-0.4cm}
    \includegraphics[width=0.5\textwidth]{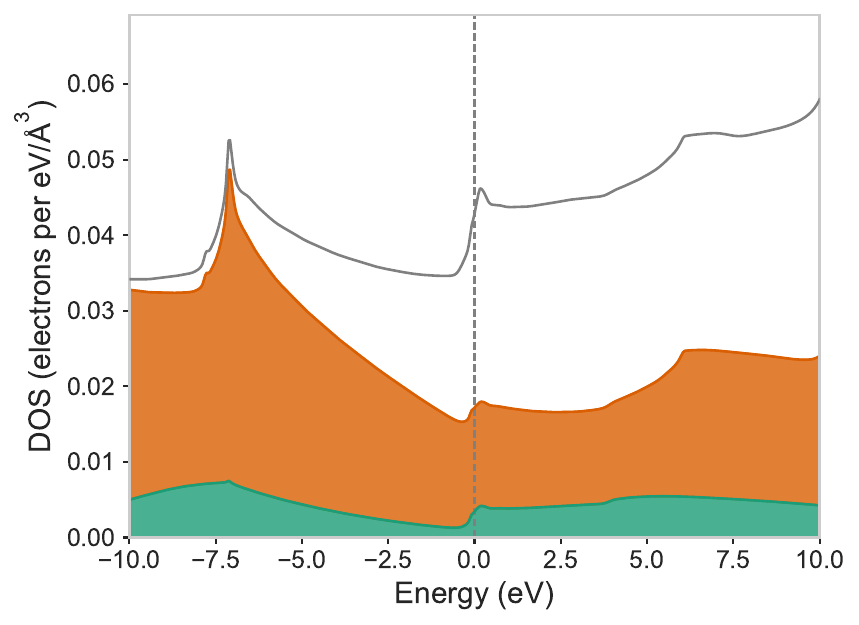}
    
    \caption{Electronic total and partial density of states (DOS) plots for the $R\bar{3}m$ phase (top) and $Pm\bar{3}m$ phase (bottom) at $375$ GPa. The projections onto the $l=0$ channel are shown in green, and onto the $l=1$ channel in orange. The Fermi level is set at $E=0$.}
    \label{elec_DOSs}
\end{figure}

Figure \ref{elec_DOSs} shows electronic density of states (DOS) plots for the trigonal and simple cubic phases at $375$ GPa using LDA, where the static-lattice enthalpy of the simple cubic phase relative to the trigonal phase is lowest.

In comparison to the trigonal phase, the $Pm\bar{3}m$ phase total DOS has a greater weight at lower energies, which lowers the total electronic energy. The Fermi energy $E_F$ is also very slightly lowered in the $Pm\bar{3}m$ phase compared to the trigonal phase. These effects become more pronounced with increasing pressure, hence the negative slope of the red line in figure \ref{energy_pv_zpe_enth}. As discussed, whilst this relative electronic energy gain is enough to overcome the contribution of the $pV$ term in a certain pressure window, it is not large enough to overcome the much larger ZPE of the simple cubic phase.

\section{\label{conclusion}Conclusions}

We have shown using a variety of XC-functionals that the high-pressure trigonal $\rightarrow$ simple cubic transition does not take place in sulfur when zero-point energies are included, and we have further shown that neither the explicit addition of core ($n=2$) electrons, nor the consideration of anharmonic corrections to the zero-point energies, are able to change this conclusion. The transition becomes even less favourable at finite temperatures.
\\\\
Whilst the SC phase becomes favourable at the static-lattice level under LDA and PW91, its comparatively much larger zero-point energy relative to the $R\bar{3}m$ phase suppresses the transition.
\\\\
As mentioned in the concluding remarks of \cite{Whaley_Baldwin}, experimental work is urgently needed to confirm these findings. The highest pressures to be considered ($\approx 500$ GPa) lie within those accessible by high-pressure diamond anvil experiments, and this pressure should also be sufficient to confirm the subsequent transition to the $Im\bar{3}m$ phase.

\section{\label{acknowledgements}Acknowledgements}

The author wishes to acknowledge Mark Johnson for useful discussions and assistance with the anharmonic calculations.

We are grateful for computational support from the UK national high performance computing service, ARCHER, for which access was obtained via the UKCP consortium and funded by EPSRC grant ref EP/P022561/1

\clearpage
\bibliography{references.bib}

\begin{thebibliography}{1}

\bibitem{Whaley_Baldwin}
J.~Whaley-Baldwin and R.~Needs, ``First-principles high pressure structure
  searching, longitudinal-transverse mode coupling and absence of simple cubic
  phase in sulfur.,'' {\em New Journal of Physics}, vol.~22, p.~023020, Mar
  2020.

\bibitem{Rudin}
S.~P. Rudin and A.~Y. Liu, ``Predicted simple-cubic phase and superconducting
  properties for compressed sulfur,'' {\em Phys. Rev. Lett.}, vol.~83,
  pp.~3049--3052, Oct 1999.

\bibitem{USPEX}
S.~S.~D. Pavel N.~Gavryushkin, Konstantin D.~Litasov and Z.~I. Popov,
  ``High--pressure phases of sulfur: Topological analysis and crystal structure
  prediction,''

\bibitem{Castep}
S.~J.~Clark, M.~Segall, C.~J.~Pickard, P.~Hasnip, M.~Probert, K.~Refson, and
  M.~C.~Payne, ``First principles methods using castep,'' {\em Zeitschrift
  f{\"u}r Kristallographie}, vol.~220, 05 2005.

\bibitem{ELK}
J.~K.~D. et~al., ``Elk code,''

\bibitem{Vinet}
P.~Vinet, J.~R. Smith, J.~Ferrante, and J.~H. Rose, ``Temperature effects on
  the universal equation of state of solids,'' {\em Phys. Rev. B}, vol.~35,
  pp.~1945--1953, Feb 1987.

\end{thebibliography}
\bibliographystyle{ieeetr}

\end{document}